\documentclass{elsart}
\usepackage{latexsym}
\usepackage{amsmath}

\begin{document}
\begin{frontmatter}
\title{Correlation Functions in Holographic Renormalization
Group Flows}

\author{Wolfgang M{\"u}ck\thanksref{email}}

\address{Dipartimento di Scienze Fisiche, Universit\`a di Napoli
``Federico II''\\ Via Cintia, 80126 Napoli, Italy}

\thanks[email]{E-mail: \texttt{mueck@na.infn.it}}

\begin{abstract}
We consider the holographic duality for a generic bulk theory of
scalars coupled to gravity. By studying the fluctuations around
Poincar\'e invariant backgrounds with non-vanishing scalars, with the
scalar and metric 
boundary conditions considered as being independent, we obtain all one-
and two-point functions in the dual renormalization group flows of the
boundary field theory. Operator and vev flows are explicitly
distinguished by means of the physical condensates. The method
is applied to the GPPZ and Coulomb branch flows, and field theoretical
expectations are confirmed. 
\end{abstract}
\end{frontmatter}

\providecommand{\e}{}
\renewcommand{\e}[1]{\mathrm{e}^{#1}}
\providecommand{\F}{\mathrm{F}}

\providecommand{\tg}{\tilde{g}}
\providecommand{\tR}{\tilde{R}}
\providecommand{\tG}{\tilde{\Gamma}}

\providecommand{\btg}{\bar{\tg}}
\providecommand{\bg}{\bar{g}}

\providecommand{\hTT}{\breve{h}}
\providecommand{\hh}{\hat{h}}
\providecommand{\rh}{\check{h}}

\providecommand{\bp}{\bar{\phi}}
\providecommand{\hp}{\hat{\varphi}}
\providecommand{\rp}{\check{\varphi}}

\section{Introduction}
\label{intro}
In the past years, holographic dualities have been used extensively in
the study of quantum field theories. According to the holographic
principle, the properties of a quantum field theory living on the boundary of a
certain bulk space-time are in one-to-one correspondence with the
dynamics of a field theory living in that bulk \cite{tHooft93,Susskind95}. 
The duality most studied  (and very well understood)
is the correspondence between conformal field
theories and field theories living on anti-de Sitter bulk spaces (AdS/CFT
correspondence), \emph{e.g.}, the duality between $\mathcal{N}=4$
SYM theory and gauged $\mathcal{N}=8$ super gravity on $AdS_5\times
S^5$ \cite{Maldacena98,Gubser98-1,Witten98-1}. 

In general, the bulk theories also allow for solutions that
describe asymptotically AdS spaces. In the case of $\mathcal{N}=8$
super gravity, these gravitational backgrounds
have been interpreted as the duals of $\mathcal{N}=4$ SYM theory either
perturbed by the insertion of relevant operators or given a non-zero
vacuum expectation value (vev) of some scalar operator
\cite{Akhmedov98,Girardello98,Khavaev00,Distler99a,Freedman99a,Freedman00b,Girardello99a,Distler99b}. Formulated more generally, asymptotically AdS bulk
geometries describe the renormalization group (RG) flows of 
boundary field theories with conformal UV fixed points
\cite{Alvarez99,Porrati99,Balasubrasmanian99,DeWolfe00b,Sahakian00,Petrini00,Porrati00a,Porrati01,Erdmenger01}. 

Using the recipe of the AdS/CFT correspondence, the correlation
functions of operators in boundary field theories with RG flows can be
obtained by 
studying the fluctuations in the bulk theory around the
background flow solutions. This procedure has been applied to the
operator flow of Girardello, Petrini, Porrati and Zaffaroni (GPPZ)
\cite{Girardello99a} as well as to some of the Coulomb branch vacuum
flows of $\mathcal{N}=4$ SYM theory \cite{Freedman00b,Brandhuber99b}. The
correlation functions obtained so far inlude various two-point
functions for scalar operators and the traceless transveral parts of
the two-point functions of the energy momentum tensor
\cite{Chepelev99-1,Giddings00,Rashkov99-3,Freedman00b,Anselmi00a,DeWolfe00a,Arutyunov00a,Bianchi00}, as well as the two-point
functions of $\mathcal{R}$-currents \cite{Brandhuber00a}.
In \cite{Bianchi00}, also the fermion,
vector and gravitino sectors of the super gravity fluctuations are
discussed.  

In the present paper, we shall consider in detail a generic bulk
theory consisting of scalars coupled to gravity. Our improvements with
respect to preceding work \cite{DeWolfe00a,Arutyunov00a,Bianchi00}
will be two-fold. First, although the fluctuations of the active scalars
(scalars having a non-zero background) are coupled to the metric
fluctuations, their respective boundary values are prescribed
independently, since they are the sources in the boundary field
theory. This enables us to obtain, in addition to the scalar two-point
function, the complete energy momentum tensor two-point function as
well as the mixed scalar--energy momentum tensor two-point
function. Second, we are able to distinguish explicitly between the
operator and the vev flows. In the former, the boundary field theory
is perturbed by inserting a scalar operator into the boundary field
theory action with a finite coupling, whereas in the latter there is a
change of the vacuum of the field theory. 
While all one-point functions in the operators flows vanish, the vev flows
explicitly exhibit a condensate. These one-point functions are
generated by the addition of counter terms necessary to render
the generating functional finite.

As a preliminary, it is instructive to consider the foundation of
holography, namely the correspondence formula
\begin{equation}
\label{correspondence}
 \left\langle \e{-\int \sqrt{\hat{g}}\, d^dx\, \mathcal{O}(x) \hat{\phi}(x)} \right\rangle_0 
 \equiv \e{-S[\hat{\phi},\hat{g}_{ij}]}~.
\end{equation}
On the right hand side, $S[\hat{\phi},\hat{g}]$ denotes the (properly
renormalized) on-shell action of the bulk field theory as a functional
of prescribed boundary data $\hat{\phi}$ and $\hat{g}_ {ij}$. On the
left hand side, we have the generating functional for the boundary
quantum field theory, which lives on a background with metric
$\hat{g}_{ij}$, and whose action functional is equal to the action
functional at an
ultraviolet fixed point perturbed by the insertion of the relevant
operator $\mathcal{O}$ with coupling $\hat{\phi}$. The expectation
value on the left 
hand side of eqn.\ \eqref{correspondence} is taken with respect to the
unperturbed fixed point action, indicated by the index 0. From eqn.\
\eqref{correspondence} follow straightforwardly the correlation
functions of the boundary field theory,
\begin{align}
\label{onepoint}
  \left\langle \mathcal{O}(x) \right\rangle_{\hat{\phi}} &= 
  \frac1{\sqrt{\hat{g}(x)}} \frac{\delta S}{\delta \hat{\phi}(x)}~,\\
\label{twopoint}
  \left\langle \mathcal{O}(x) \mathcal{O}(y) \right\rangle_{\hat{\phi}} &= 
  \frac1{\sqrt{\hat{g}(x)}\sqrt{\hat{g}(y)}} \left[ 
  - \frac{\delta^2 S}{\delta \hat{\phi}(x)\delta \hat{\phi}(y)}  
  + \frac{\delta S}{\delta \hat{\phi}(x)} 
  \frac{\delta S}{\delta \hat{\phi}(y)}\right]~, \quad \text{etc.}
\end{align}
In eqns.\ \eqref{onepoint} and \eqref{twopoint}, the expectation
values are calculated in the perturbed boundary field
theory, indicated by the index $\hat{\phi}$. 

In the remainder of this section, ley us give a brief outlook of the
rest of the paper. We shall explain 
the general method in Sec.\ \ref{method}, which is subdivided into
two parts. In subsection \ref{eqmot} we present in detail the equations of
motion and the on-shell action of the bulk field theory, with special
emphasis on our new gauge choice. 
The resulting equations of motion for the active scalars and the bulk
graviton are coupled, so that, in the case of one scalar, they give
rise to a third order ordinary differential equation (ODE). However,
one solution is 
found, which guarantees the reduction of the problem to a second order
ODE. The reader is referred to the appendix for
matters of notation and for some relevant formulae, which are
used in subsection \ref{eqmot}.
In subsection \ref{sourceresponse} we
explain how the on-shell action is expressed in terms of the
sources of the boundary field theory, which is essential for obtaining
the correlation functions. Three cases are distinguished: operator
flows involving active scalar operators of dimension $\Delta>d/2$, vev
flows for these operators, and vev flows for operators of dimension
$\Delta= d/2$. In the latter two, the appearance of the
condensate is demonstrated.
In Sec.\ \ref{flows} we shall apply our method to the GPPZ and Coulomb
branch flows. The algebra involved, especially the factorization of
the third order differential equations, is quite complex. Therefore,
we had to make extensive use of symbolic algebra software (Maple). The
reader is guided through the essential steps of the calculation with
only the results of them given. Finally, Sec.\ \ref{conc} contains
conclusions.

\section{General Method}
\label{method}
\subsection{Equations of Motion and on-Shell Action}
\label{eqmot}
We consider a generic bulk field theory containing scalars coupled to gravity
with the following action, 
\begin{equation}
\label{S}
  S = S_{\mathrm{bulk}} + S_{\mathrm{surf}}~,
\end{equation}
where the bulk term is given by\footnote{Our conventions for
the curvature tensor are 
$R^\mu{}_{\nu\rho\lambda}=\partial_\rho \Gamma^\mu{}_{\nu\lambda} +
\Gamma^\mu{}_{\rho\sigma}\Gamma^\sigma{}_{\nu\lambda}
-(\rho\leftrightarrow\lambda)$, $R_{\mu\nu}=R^\rho{}_{\mu\rho\nu}$.
We have adorned with a tilde quantities belonging to $(d+1)$ space
to distinguish them from those of the $d$-dimensional
hyper surfaces.} 
\begin{equation}
\label{Svol}
  S_{\mathrm{bulk}} =  \int d^{d+1} x \sqrt{\tg} \left[ -\tR +\frac12
  \tg^{\mu\nu} \partial_\mu \phi^I \partial_\nu \phi^I -V(\phi)
  \right]~.
\end{equation}
We assume that the potential $V(\phi)$ can be written
in terms of a ``super potential'' $U(\phi)$ as 
\begin{equation}
\label{V}
  V = \frac{d}{4(d-1)} U^2 -\frac12 
  \left( \frac{\partial U}{\partial \phi^I} \right)^2~.
\end{equation}
This relation holds for actions $S$ stemming from super gravity
theories, but might in fact be more general (see \cite{Skenderis99} and
references therein).   
The surface term of the action, $S_{\mathrm{surf}}$, consists of  the
Gibbons--Hawking term containing the second fundamental form, $H$,  
and a series of counter terms
\cite{Henningson98-2,Liu98-1,Arutyunov98-1,deHaro00a}, of which we shall
write explicitly only the first one, since the others do not
contribute to the non-local correlation functions,
\begin{equation}
\label{Ssurf}
  S_{\mathrm{surf}} =  \int d^d x \sqrt{g} \left[ 2 H - U(\phi) +\cdots
  \right]~.
\end{equation}
Our specific choice for the counter term is inspired by the
Hamilton-Jacobi method in holography \cite{deBoer00a}.

We are interested in the on-shell action for fluctuations around a
given background solution. Let us write 
\[ \phi^I \to \phi^I + \varphi^I~, \quad \tg_{\mu\nu} \to
\tg_{\mu\nu} + h_{\mu\nu}~,\]
where $\varphi^I$ and $h_{\mu\nu}$ are fluctuations around a 
solution given by $\phi^I$ and $\tg_{\mu\nu}$. Then, the first order
variation of $S_{\mathrm{bulk}}$ is obtained as ($N^\mu$ is the surface
normal vector)  
\begin{equation}
\label{dSvol}
  \delta S_{\mathrm{bulk}}[\phi,\tg;\varphi, h] = 
  \int d^d x \sqrt{g} \left[ N_\mu \tilde{\nabla}_\nu h_{\lambda\rho}
  \left( \tg^{\mu\nu}\tg^{\lambda\rho} - \tg^{\mu\lambda}\tg^{\nu\rho}
  \right) + N^\mu \varphi^I \partial_\mu
  \phi^I \right]~,
\end{equation}
where we have used the equations of motion,
\begin{gather}
\label{eqmotphi}
  \tilde{\nabla}^2 \phi^I + \frac{\partial V}{\partial \phi^I} =0~,\\
\label{eqmoteinst}
  \tR_{\mu\nu} -\frac12 \partial_\mu \phi^I \partial_\nu \phi^I
  -\frac12 \tg_{\mu\nu} \left[ \tR -\frac12 \tg^{\lambda\rho}
  \partial_\lambda \phi^I \partial_\rho \phi^I +V(\phi) \right] =0~,
\end{gather}
in order to eliminate the bulk integral.

To eqn.\ \eqref{dSvol}, we add the first order variation of the
surface term, $S_{\mathrm{surf}}$, and find 
\begin{equation}
\label{dS}
  \delta S[\phi,\tg;\varphi,h] =  \int d^d x \sqrt{g} \left[ 
  \left( \frac{\partial_r \phi^I - n^i \partial_i \phi^I}n 
  - \frac{\partial U}{\partial \phi^I} \right) \varphi^I 
  - \left( H^{ij} - H g^{ij} +\frac12 U g^{ij} \right) h_{ij}
  \right]~.
\end{equation}
In order to obtain eqn.\ \eqref{dS}, we have made use of the geometric
relations governing hyper surfaces, which are listed in
appendix~\ref{geom}. 

Eqn.\ \eqref{dS} can be used to calculate the on-shell action to
second order in the fluctuations around a given background solution,
$\bp$, $\btg$, using the simple formula
\begin{equation}
\label{dS2}
  S[\bp+\varphi,\btg+h] = S[\bp,\btg] + \delta
  S[\bp+\varphi/2,\btg+h/2; \varphi, h] + \mathcal{O}((\varphi,h)^3)~.
\end{equation} 

Specifically, we are interested in $d$-Poincar\'e invariant
background solutions of the form
\begin{equation}
\label{bgansatz}
  \bp^I = \bp^I(r)~, \quad 
  ds^2 = dr^2 +\e{2A(r)} \eta_{ij} dx^i dx^j~. 
\end{equation}
For a potential $V$ of the form \eqref{V}, such an ansatz satisfies
the equations of motion \eqref{eqmotphi} and \eqref{eqmoteinst}, if
\cite{Freedman99a,Skenderis99,DeWolfe00b} 
\begin{equation}
\label{bg}
  \partial_r \bp^I{} = \frac{\partial U}{\partial \bp^I}~, \quad 
  \partial_r A(r) = -\frac1{2(d-1)} U[\bp(r)]~.
\end{equation}

Let us continue with linearizing the equations of motion, eqns.\
\eqref{eqmotphi} and \eqref{eqmoteinst}, around the background
\eqref{bgansatz}. First, eqn.\ \eqref{eqmotphi} yields
\begin{equation}
\label{phieqn}
  \left( \partial^2_r +dA'\partial_r +\e{-2A} \Box \right) \varphi^I
  +\frac{\partial^2 V}{\partial \bp^I \partial \bp^J} \varphi^J
  +\frac{\partial V}{\partial \bp^I} h_{rr} 
  -\frac12 \frac{\partial U}{\partial \bp^I} \left( \partial_r h_{rr}
  -\partial_r h^i_i +2 \partial_i h^i_r \right) =0~.
\end{equation}
Regarding our notation, $\Box=\partial^i\partial_i$ and $\partial^i
=\eta^{ij}\partial_j$, but $h^i_r= \bg^{ij} h_{jr}$ and $h^i_j =
\bg^{ik} h_{kj}$. 

Second, multiplying eqn.\ \eqref{eqmoteinst} by $N^\mu N^\nu$ and
using the equation of Gauss, eqn.\ \eqref{geom:gauss2}, one finds 
\begin{equation}
\label{einstnormal}
  R +H^i_j H^j_i -H^2 
  + N^\mu N^\nu \partial_\mu \phi^I \partial_\nu \phi^I 
  - \frac12 \tg^{\mu\nu} \partial_\mu \phi^I \partial_\nu \phi^I 
  +V(\phi)=0~,
\end{equation}
whose linear form is
\begin{equation}
\label{normaleqn}
  \e{-2A} P^i_j h^j_i 
  +\frac12 U(\bp) (\partial_r h^i_i -2 \partial_i h^i_r )
  + V(\bp) h_{rr} + \frac{\partial U}{\partial \bp^I} \partial_r \varphi^I 
  + \frac{\partial V}{\partial \bp^I} \varphi^I =0~.
\end{equation}
The symbol $P^i_j$ denotes the transversal projector,
\begin{equation}
\label{P}
  P^i_j = \partial^i \partial_j - \delta^i_j \Box~.
\end{equation}

Third, the mixed components of Einstein's equation are obtained by
multiplying eqn.\ \eqref{eqmoteinst} by $N^\mu \delta^\nu_i$ and using the
equation of Codazzi, eqn.\ \eqref{geom:codazzi}, with the result
\begin{equation}
\label{einstmixed}
  \partial_i H - \nabla_j H^j_i -\frac12 N^\mu \partial_\mu \phi^I
  \partial_i \phi^I =0~.
\end{equation}
Linearizing eqn.\ \eqref{einstmixed} yields 
\begin{equation}
\label{mixedeqn}
  \partial_r (\partial^j h_j^i -\partial^i h^j_j) + P^i_j h^j_r
  -\frac12 U(\bp) \partial^i h_{rr} 
  - \frac{\partial U}{\partial \bp^I} \partial^i \varphi^I =0~.
\end{equation}

Last, the only part of the tangential components of eqn.\
\eqref{eqmoteinst} that is independent of eqns.\ \eqref{phieqn},
\eqref{normaleqn} and \eqref{mixedeqn} is the following equation for the
traceless transversal part of $h^i_j$, 
\begin{equation} 
\label{tangeqn}
  \left( \partial^2_r +dA' \partial_r +\e{-2A}\Box \right) \hTT^i_j =0~,
\end{equation}
where
\begin{equation}
\label{hTT}
  \hTT^i_j = \frac1{\Box^2} \Pi^i_j{}^k_l h^l_k~, \quad
  \Pi^i_j{}^k_l = \frac12 ( P^{ik} P_{jl} + P^i_l P^k_j ) 
  -\frac1{d-1} P^i_j P^k_l~.
\end{equation}
Eqn.\ \eqref{tangeqn} is identical to the equation for a free massless
scalar, as is well known. 

After differentiating eqn.\ \eqref{normaleqn} with respect to
$r$, one can eliminate $P^i_j h^j_i$ using eqns.\ \eqref{normaleqn} and
\eqref{mixedeqn}, and then use eqn.\ \eqref{phieqn} to obtain 
\begin{multline}
\label{hphieqn}
  \frac{d}2 U(\bp)\partial_r h_{rr} -2 V(\bp) h_{rr} +(d-1) \e{-2A} \Box h_{rr}
  +2(d-1) \frac{\partial U}{\partial \bp^I} \partial_r \varphi^I 
  - 2 \frac{\partial V}{\partial \bp^I} \varphi^I \\
  +(d-1) \partial_r (\partial_r h^i_i -2 \partial_i h^i_r)
  - U(\bp) (\partial_r h^i_i -2 \partial_i h^i_r) =0~.
\end{multline}
The two equations \eqref{phieqn} and \eqref{hphieqn} depend on the
three variables $\varphi^I$, $h_{rr}$, and $(\partial_r h^i_i -2
\partial_i h^i_r)$, the last of which can be eliminated by fixing a
suitable gauge. We shall do this now.

Under a diffeomorphism, $\delta \tg_{\mu\nu} =
\bar{\tilde{\nabla}}_\mu \xi_\nu + \bar{\tilde{\nabla}}_\nu \xi_\mu$, the
components of the metric fluctuations transform as 
\begin{align}
\label{hijtrans}
  \delta h_{ij} &= \partial_i \xi_j + \partial_j \xi_i +2A'
  \bg_{ij} \xi_r~, \\
\label{hirtrans}
  \delta h_{ir} &= \partial_i \xi_r +\partial_r \xi_i - 2A'
  \xi_i~,\\ 
\label{hrrtrans}
  \delta h_{rr} &= 2 \partial_r \xi_r~.
\end{align}
Thus, the combination $\partial_r h^i_i -2 \partial_i h^i_r$, which
appears in the equations of motion \eqref{phieqn} and
\eqref{normaleqn}, transforms as 
\begin{equation}
\label{trans1}
  \delta(\partial_r h^i_i -2 \partial_i h^i_r) = 2d\, \partial_r (A'
  \xi_r) -2 \e{-2A} \Box \xi_r~.
\end{equation}
Hence, we can use first $\xi_r$ to fix $\partial_r h^i_i -2 \partial_i
h^i_r =0$, and then $\xi_i$ to obtain $h^i_r=0$. Thus, our gauge is
\begin{equation}
\label{gauge}
  \partial_r h^i_i = h^i_r = 0~.
\end{equation}
The gauge \eqref{gauge} is accessible via non-singular gauge
transformations and does not involve the field sources.\footnote{In
contrast, the gauge $\partial_r h^i_i -2 \partial_i h^i_r =
dA'h_{rr}$, which would eliminate the terms with $\partial_r h_{rr}$
from eqn.\ \eqref{hphieqn}, needs a singular $\xi_r$ in the
asymptotic region ($r\to\infty$). The fields would thus aquire a
spurious asymptotic behaviour, and we must reject this gauge. 
One could also use $\xi_r$ to eliminate $h_{rr}$, as was done in
\cite{DeWolfe00a}. However, the resulting fluctuation equations
appeared to be difficult to analyze.} 

In the gauge \eqref{gauge}, the equations of motion \eqref{phieqn},
\eqref{normaleqn} and \eqref{mixedeqn} become 
\begin{gather}
\label{phieqn2}
  \left( \partial^2_r +dA'\partial_r +\e{-2A} \Box \right) \varphi^I
  +\frac{\partial^2 V}{\partial \bp^I \partial \bp^J} \varphi^J
  +\frac{\partial V}{\partial \bp^I} h_{rr} 
  -\frac12 \frac{\partial U}{\partial \bp^I} \partial_r h_{rr}
  =0~,\\
\label{normaleqn2}
  \e{-2A} P^i_j h^j_i 
  + V(\bp) h_{rr} + \frac{\partial U}{\partial \bp^I} \partial_r \varphi^I 
  + \frac{\partial V}{\partial \bp^I} \varphi^I =0~,\\
\intertext{and} 
\label{mixedeqn2}
  \partial_r \partial^j h_j^i -\frac12 U(\bp) \partial^i h_{rr} 
  - \frac{\partial U}{\partial \bp^I} \partial^i \varphi^I =0~,
\end{gather}
respectively, whereas eqn.\ \eqref{hphieqn} becomes
\begin{equation}
\label{hphieqn2}
  \frac{d}2 U(\bp)\partial_r h_{rr} -2 V(\bp) h_{rr} +(d-1) \e{-2A} \Box h_{rr}
  +2(d-1) \frac{\partial U}{\partial \bp^I} \partial_r \varphi^I 
  - 2 \frac{\partial V}{\partial \bp^I} \varphi^I =0~.
\end{equation}
Eqns.\ \eqref{phieqn2} and \eqref{hphieqn2} form a system of coupled
ODEs for $h_{rr}$ and $\varphi^I$. One solution can
be easily found by trial and error and is given by
\begin{equation}
\label{sol1}
  \varphi^I = \chi \frac{\partial U}{\partial \bp^I}~,\quad
  h_{rr} = 2 \partial_r \chi~,
\end{equation}
where $\chi$ satisfies the first order ODE
\begin{equation}
\label{chieqn}
  U(\bp) \partial_r \chi + \partial_r U[\bp(r)] \chi +\frac{2(d-1)}d
  \e{-2A} \Box \chi=0~.
\end{equation}
Hence, in the case of only one active scalar, $\varphi$, the
ODE \eqref{chieqn} directly translates into a first
order ODE of the form $[\partial_r +c(r)]\varphi =0$ for the special
solution given by eqn.\ \eqref{sol1}, with a unique function
$c(r)$. This means that the third order equation of motion for the
general $\varphi$ stemming from the system of equations \eqref{phieqn2} and
\eqref{hphieqn2} factorizes into the form 
\begin{equation}
\label{eqnfactor}
 [ \partial^2_r +a(r) \partial_r +b(r)][\partial_r +c(r)]\varphi
 =0~.
\end{equation}
Finding the functions $a(r)$ and $b(r)$ is straightforward and is
similar to the division of polynomials. This fact makes our approach
feasible.

We shall now turn our attention to the on-shell action. It is
straightforward to 
realize that, for a background of the form \eqref{bgansatz}, both the
background value and the first order variation of $S$ vanish,
\begin{align}
\label{Sbg}
  S[\bp,\btg] &= 0~,\\
\label{dSbg}
  \delta S[\bp,\btg;\varphi,h] &= 0~.
\end{align}
Eqn.\ \eqref{Sbg} means that the normalization of the expectation
values in eqns.\ \eqref{onepoint} and \eqref{twopoint} is the same as
for the unperturbed theory. Moreover, it is necessary in supersymmetric
settings. Eqn.\ \eqref{dSbg} seems to imply that holographic one-point
functions vanish. This observation has already been made by Arutyunov, Frolov
and Theisen \cite{Arutyunov00a}. However, this fact survives only, if no
additional counter terms are needed, which is the case for operator
flows. We shall discuss this point in subsection \ref{sourceresponse}.

In order to obtain the term of $S$ quadratic in the fluctuations, we
expand the second term of the right hand side of eqn.\ \eqref{dS2} and
obtain, after substituting the background \eqref{bgansatz},
\begin{equation}
\label{S2}
\begin{split}
  S[\bp+\varphi,\btg+h] &=  \int d^dx \sqrt{\bg} \left[ \frac12
  \varphi^I \partial_r \varphi^I - \frac12
  \frac{\partial^2 U}{\partial\bp^I \partial\bp^J} \varphi^I \varphi^J
  +\frac14 h^i_j \partial_r h^j_i - \frac14 h^i_i \partial_r h^j_j
  \right. \\
  &\quad \left. -\frac14 \frac{\partial U}{\partial \bp^I} \varphi^I
  (h_{rr} +h^i_i) -\frac18 U(\bp) h^i_i h_{rr} -\frac12 h^i_j
  \partial_i h^j_r + \frac12 h^i_i \partial_j h^j_r \right]+\cdots~.
\end{split}
\end{equation}
It takes only a few
steps to show that, by virtue of eqn.\ \eqref{mixedeqn} and the gauge
\eqref{gauge}, eqn.\ \eqref{S2} can be rewritten as 
\begin{equation}
\label{S3}
\begin{split}
  S &= \frac14 \int d^d x \sqrt{\bg} \left[ 
  2 \varphi^I \partial_r \varphi^I 
  - 2 \frac{\partial^2 U}{\partial\bp^I\bp^J} \varphi^I \varphi^J 
  - \frac{\partial U}{\partial\bp^I} \varphi^I h_{rr}  \right. \\
  &\quad \left.
  +h^i_j \frac1{\Box^2} \partial_r \left(\Pi^j_i{}^k_l h^l_k\right)
  + \frac{d}{d-1} h^i_j P^j_i \frac1{\Box^2} \partial_r (P^k_l h^l_k)
  \right]~.
\end{split}
\end{equation}
The last term on the first line will not contribute to the non-local
two-point functions.

\subsection{Sources and Responses}
\label{sourceresponse}
We shall begin this section introducing the reader to what we exactly
mean by the terms ``sources'' and ``responses.'' In the case of scalar
fields in a pure AdS background, these notions are identical to the
regular and irregular boundary data, respectively
\cite{Breitenlohner82}. In the AdS/CFT correspondence, they have been
used in \cite{Klebanov99-1,Mueck99-4}. 
For simplicity, let us consider a single scalar field 
that couples to an operator of scaling dimension $\Delta$. 
There are three different cases, operator flows involving 
scalar operators of dimension $\Delta>d/2$, vev flows involving these
scalars, and vev flows involving operators of dimension $\Delta=d/2$.
We shall start by analyzing in detail the operator
flows, and consider the differences in the other two cases later. 

The behaviour of the fields $\varphi$ and $h^i_j$ for large $r$ can be
described as \cite{Witten98-1,Henningson98-2,deHaro00a}
\begin{align}
\label{phiasym}
  \varphi(x,r) &=\e{(\lambda-d)r/l} \left[ \hp(x) +\cdots\right] 
  + \e{-\lambda r/l} \left[\rp(x) + \cdots\right]~,\\
\label{hasym}
  h^i_j (x,r) &= \left[\hh^i_j(x) + \cdots \right]+ 
  \e{-d r/l} \left[\rh^i_j(x) + \cdots\right]~.
\end{align}
Here, $\lambda>d/2$, so that the first pairs of brackets on the right hand
sides of eqns.\ \eqref{phiasym} and \eqref{hasym} denote the leading
series. The coefficcients $\hp(x)$ and $\hh^i_j(x)$ are regarded as
the sources, and the remaining terms in the brackets depend locally on
them.\footnote{There is another way of defining the sources, namely by
imposing Dirichlet boundary conditions on the fields at some cut-off
boundary. The boundary values are the bare sources, which have to be
rescaled when sending the cut-off to infinity. This method has been
used successfully in the AdS/CFT correspondence, but it is technically very
difficult to solve a Dirichlet boundary value problem in the RG flow 
backgrounds considered here.} 
The second pairs of brackets denote the subleading series,
which are determined by the responses $\rp(x)$ and
$\rh^i_j(x)$.\footnote{Generically, the
source and response series are in powers of $\e{-2r/l}$ with
coefficients that depend locally on the sources and responses,
respectively. However, for $\lambda=d/2+k$, with 
positive integer $k$, as well as for even $d$, the source series
contain terms of the form $r/l\,\e{-2nr/l}$ with $n\ge k$ and $n\ge d/2$,
respectively.}  
The general solution to the coupled equations of motion, eqns.\ \eqref{phieqn2}
and \eqref{hphieqn2}, is determined uniquely by specifying three
integration constants, say, $\hp$, $P^i_j\hh^j_i$ and $\rp$, but it is
in general not regular in the bulk. Thus, imposing one suitable
regularity condition, only two constants need to be specified, which
we choose to be the sources, $\hp$ and $P^i_j\hh^j_i$. Then, the responses,
$\rp$ and $P^i_j\rh^j_i$, are uniquely determined and give rise to the
holographic two-point functions. It seems clear that, since only one
condition can be imposed, one cannot demand regularity for both
fields, $\varphi$ and $h^i_j$. 
Notice that, in general, each of the responses will depend on
both sources. This is not so in the pure AdS background, where the
scalar and gravity equations of motion decouple. 

In order to analyze the equations of motion in the asymptotically AdS
region of the bulk (for large $r$), we let $A(r)=r/l+\cdots$. 
Moreover, we need the
following expressions of the potentials $U$ and $V$ close to the UV
fixed point \cite{Kalkkinen01b},
\begin{align}
\label{Uasym} 
  U(\phi) &= -\frac{2(d-1)}l + \frac{\lambda-d}{2l} \phi^2 + \cdots~,\\
\label{Vasym}
  V(\phi) &= \frac{d(d-1)}{l^2} - \frac12 m^2 \phi^2 + \cdots~, 
\end{align}
where $m^2 l^2 = \lambda(\lambda-d)$. Since the background field
satisfies eqn.\ \eqref{bg}, eqn.\ \eqref{Uasym} tells us that $\bp$
behaves as 
\begin{equation}
\label{bpasym}
 \bp(r) = \hat{\bp}\,\e{(\lambda-d)r/l} +\cdots~.
\end{equation} 

Using eqn.\ \eqref{normaleqn2} one can determine the asymptotic
behaviour of $h_{rr}$,
\begin{equation}
\label{hrrasym}
  h_{rr}(x,r) = -\frac{l^2 \e{-2r/l}}{d(d-1)} P^i_j \hh^j_i +
  \frac{\lambda-d}{d-1} \hat{\bp} \hp\, \e{2(\lambda-d)r/l} +\cdots 
  + \frac{2m^2l^2}{d(d-1)} \hat{\bp} \rp\, \e{-dr/l} +\cdots~. 
\end{equation}
Then, from eqn.\ \eqref{mixedeqn2} one finds the asymptotic behaviour
of $P^i_j h^j_i$, 
\begin{equation}
\label{Phasym}
  P^i_j h^j_i(x,r) = P^i_j \hh^j_i(x) - \frac{l^2}{2d} \e{-2r/l} \Box
  P^i_j \hh^j_i +\cdots +
  \frac{(\lambda-d)(2\lambda-d)}{d^2} \hat{\bp} \Box \rp \,\e{-dr/l} +\cdots~.
\end{equation}
It is reassuring that the term with $\e{2(\lambda-d)r/l}$ vanishes, and
that one reproduces the behaviour of eqn.\ \eqref{hasym}. 

The traceless transversal part of $h^i_j$ obeys the second order
ODE \eqref{tangeqn}. The asymptotic behaviour of the solution is
given by  
\begin{equation}
\label{httasym}
  \hTT^i_j(x,r) = \hat{\hTT}^i_j(x) +
  \frac{l^2}{2(d-2)} \e{-2r/l} \Box  \hat{\hTT}^i_j(x) +\cdots 
  + \check{\hTT}^l_k(x) \e{-dr/l} +\cdots~.
\end{equation}
As in the pure AdS background, the response $\check{\hTT}^l_k$ is
determined in terms of the source $\hat{\hTT}^i_j$ by imposing the
condition that the solution be regular in the bulk.  

Let us now return to the on-shell action, eqn.\ \eqref{S3}. After
inserting the expansions \eqref{phiasym}, \eqref{Phasym} and
\eqref{httasym}, the finite terms give us the generating funcional for
two-point functions of the boundary field theory. However, we must
first look at the divergent terms. The leading divergent terms
(proportional to $\e{dr/l}$ and $\e{(2\lambda-d)r/l}$) should vanish,
since the appropriate counter terms have already been added. 
In eqn.\ \eqref{S3}, there are no terms proportional to $\e{dr/l}$,
while it is easy to see that the terms proportional to
$\e{(2\lambda-d)r/l} \hp^2$ cancel.
Moreover, the remaining divergent terms must be local, so that they
can be cancelled by adding further local counter terms (involving the
curvature and derivatives of $\phi$). The two terms on the second line
of eqn.\ \eqref{S3} might worry us, since they are non-local in the
sources. However, from eqns.\ \eqref{Phasym} and \eqref{httasym}
follows
\begin{multline}
\label{divcancel}
  h^i_j \frac1{\Box^2} \partial_r \left(\Pi^j_i{}^k_l h^l_k\right)  
  + \frac{d}{d-1} h^i_j P^j_i \frac1{\Box^2} \partial_r (P^k_l h^l_k) \\
  = -\frac{l}{d-2} \hh^i_j \frac{\Pi^j_i{}^k_l}{\Box} \hh^l_k
  \,\e{-2r/l} + \frac{l}{d-1} \hh^i_j \frac{P^j_iP^k_l}{\Box} \hh^l_k
  \, \e{-2r/l} +\cdots~,
\end{multline}
and the non-local terms (stemming from the terms of the
projectors with four distinct indices) cancel. Hence, the remaining
divergent terms can be cancelled by adding appropriate local counter
terms. These counter terms will contribute further contact terms to
the correlators, but we will not consider them explicitly. 

We can now turn our attention to the finite, non-local part of the
action. Inserting eqns.\ \eqref{phiasym}, \eqref{Phasym} and
\eqref{httasym} into eqn.\ \eqref{S3}, we obtain
\begin{equation}
\label{Sfin}
  S_{\mathrm{fin}} = \frac1l \int d^dx \left[ -\left(\lambda-\frac{d}2
  \right) \hp\rp - \frac{d}4 \hh^i_j \frac{\Pi^j_i{}^k_l}{\Box^2}
  \rh^l_k - \frac{d^2}{4(d-1)} \hh^i_j \frac{P^j_i P^k_l}{\Box^2}
  \rh^l_k \right]~.
\end{equation}
From $S_{\mathrm{fin}}$, correlation functions are obtained using the
correspondence formula \eqref{correspondence}, as in eqns.\
\eqref{onepoint} and \eqref{twopoint}. When doing so, we shall multiply
$S_{\mathrm{fin}}$ by a constant of dimension length, $\kappa$, in order
to make the exponent dimensionless. 

The first term in eqn.\ \eqref{Sfin} is responsible for the scalar
two-point function, $\langle \mathcal{O} \mathcal{O}\rangle$. From the
AdS/CFT correspondence it is known that the boundary field theory
operators have dimension $\Delta=\lambda>d/2$ at the conformal fixed
point. The second and third terms give the two-point function of the
energy momentum tensor, $\langle T^i_j T^k_l \rangle$. However, since
the scalar $\varphi$ couples to $P^i_j h^j_i$, the response $\rp$ will
also depend on the source $\hh^i_j$, and, \emph{vice versa}, the
response $P^i_j \rh^j_i$ will depend on the source $\hp$. Thus, the
first and third terms of eqn.\ \eqref{Sfin} give also rise to a two-point
function of the form $\langle T^i_j \mathcal{O} \rangle$. This
two-point function satisfies the Ward identity for $4d$ Poincar\'e
invariance, 
\begin{equation}
\label{Ward}
  \frac{\partial}{\partial x^i} \left\langle T^i_j (x) \mathcal{O}(y)
  \right\rangle =0~,
\end{equation}
since it is proportional to the transversal projector $P^i_j$.

Let us consider now the case $\lambda<d/2$, which corresponds to
vev flows. The formulae
\eqref{phiasym} to \eqref{httasym} remain valid as before, but the
interpretation of the fields $\hp$ and $\rp$ changes. In fact, since
$\lambda-d<-\lambda$, the series with $\rp$ in eqn.\
\eqref{phiasym} becomes the leading one. Correspondincly, we interpret
$\rp$ as the source, whereas $\hp$ is the response. 
The background, behaving as in
eqn.\ \eqref{bpasym}, follows the subleading behaviour of the
fluctuations. This means that the background source is zero,
\emph{i.e.}, there is no operator insertion in the boundary field
theory action. 

The on-shell action, eqn.\ \eqref{S3}, becomes
\begin{equation}
\label{vevS}
  S = - \frac1l \left( \lambda -\frac{d}2 \right) \int d^dx \left( \hp
  \rp + \rp^2 \e{(d-2\lambda)r/l} +\cdots \right)~,
\end{equation}
where we have not written those terms that remain unchanged with
respect to eqn. \eqref{Sfin}. However, since we now have $d>2\lambda$,
the second term is divergent (in this case, the counter term with
$U(\phi)$ cancels the sub-leading term), and we need to add a counter term,
\begin{equation}
\label{vevSct}
  S_{\mathrm{ct}} =  \frac1l \left( \lambda -\frac{d}2 \right) \int
  d^dx \sqrt{g}\, \phi^2~.
\end{equation}
Notice that the counter term must be expressed in terms of the full
fields, because of the universality of the action. Adding eqn.\
\eqref{vevS} to \eqref{vevSct}, we find the finite on-shell action,
\begin{equation}
\label{vevSfin}
  S_{\mathrm{fin}} =  \frac1l\left( \lambda -\frac{d}2 \right) \int
  d^dx \left( \rp \hp + 2 \rp\hat{\bp} +h^i_i \hat{\bp}\rp +\cdots
  \right)~. 
\end{equation}
Eqn.\ \eqref{vevSfin} gives correlation functions of operators, whose
dimension at the conformal fixed point is $\Delta=d-\lambda>d/2$. In
fact, for the AdS background, eqns.\ \eqref{Sfin} and \eqref{vevSfin}
are identical, when $\lambda$ is expressed in terms of $\Delta$. 
For RG flow backgrounds, we explicitly obtain the
condensate 
\begin{equation}
\label{vevO}
  \langle \mathcal{O}(x) \rangle = (d-2\Delta) \frac{\kappa}l \hat{\bp}~.
\end{equation}
The contact contribution to $\left\langle T^i_j \mathcal{O}\right\rangle$ from
the third term on the right hand side of eqn.\ \eqref{vevSfin} has its
origin in the fact that $T^i_j$ is the energy-momentum operator of the
boundary field theory including the source term \cite{Bianchi01}. 

Let us finally consider the special case of operators with scaling
dimension $\lambda=d/2$, which also corresponds to vev flows. 
While the background field behaves asymptotically as 
\begin{equation}
\label{d2bgasy}
  \bp(r) = \e{-dr/(2l)} \check{\bp} +\cdots~,
\end{equation}
the leading behaviour of the perturbation $\varphi$ is 
\begin{equation}
\label{d2asy}
  \varphi(r,x) = \e{-dr/(2l)} \left[\frac{r}{l} \hp(x) + \rp(x)
  \right] + \cdots~. 
\end{equation}
Again, the background behaves as the subleading part of the
fluctuation. Hence, the background source is zero, and the RG flow must be
a vev flow. 

Inserting eqn.\ \eqref{d2asy} into the action \eqref{S3}, one finds 
\begin{equation}
\label{d2S}
  S= \frac1{2l} \int d^dx \left( \frac{r}{l} \hp^2 + \hp\rp +\cdots
  \right)~,
\end{equation}
where we have not written those terms that remain unchanged with
respect to eqn. \eqref{Sfin}. The term with $r/l$ in $S$ is 
logarithmically divergent and must be cancelled by adding a suitable counter
term,\footnote{``Logarithmically'' because $r/l$ is proportional to
the logarithm of the volume of the boundary space.}
\begin{equation}
\label{Sct}
  S_{\mathrm{ct}} = -\frac1{2r} \int d^dx \sqrt{g}\, \phi^2~.
\end{equation}
As before, the counter term must be expressed in terms of the full
fields $\phi$ and $g_{ij}$, not only in terms of the fluctuations
$\varphi$ and $h^i_j$, since it should be valid for any background
solution. Adding $S_{\mathrm{ct}}$ to eqn.\ \eqref{d2S} yields the
finite action
\begin{equation}
\label{d2Sfin}
  S_{\mathrm{fin}} = -\frac1l \int d^dx \left( \frac12 \hp \rp + \hp
  \check{\bp} +\frac12 \hp \check{\bp} \hh^i_i +\cdots \right)~,
\end{equation}
where the ellipses indicate the last two terms of eqn.\
\eqref{Sfin}, which remain unchanged. One can check easily that the
formula \eqref{d2Sfin} yields the correct (\emph{i.e.}, postive definite) 
conformal two-point function $\langle\mathcal{O}(x) \mathcal{O}(y)\rangle$
in the case of the AdS background. The scalar operators have dimension
$\Delta=d/2$ at the conformal fixed point. 

The second term in eqn.\ \eqref{d2Sfin} yields the condensate of the
boundary field theory,
\begin{equation}
\label{condensate}
  \left\langle \mathcal{O}(x) \right\rangle = - \frac{\kappa}l \check{\bp}~,
\end{equation}
whereas the third term again gives a local contribution to the correlator 
$\left\langle T^i_j \mathcal{O}\right\rangle$.

\section{Particular Flow Solutions}
\label{flows}
\subsection{GPPZ Flow}
\label{GPPZ}
An explicit flow solution describing the renormalization group flow
from $\mathcal{N}=4$ SYM theory in the UV to $\mathcal{N}=1$ SYM
theory in the IR ($d=4$) was found by GPPZ \cite{Girardello98}. The
flow is generated by perturbing the SYM theory action by inserting an
operator of dimension $\Delta=3$.
The super potential $U(\phi)$ responsible for the GPPZ flow is, in our
notation (only the non-zero active scalar is shown),
\begin{equation}
\label{GPPZ:U}
  U(\phi) = -\frac3l \left( \cosh\frac{\phi}{\sqrt{3}} +1 \right)~,
\end{equation}
and the background flow solution is given by
\begin{align}
\label{GPPZ:pb}
  \bp(r) &= \sqrt{3}\, \ln \frac{1+\e{-r/l}}{1-\e{-r/l}}~, \\
\label{GPPZ:A}
  A(r) &= \frac12 \left[ \frac{r}l + \ln \left(\e{r/l}-\e{-r/l} \right)
  \right]~.
\end{align} 
Towards the horizon, the background field $\bp$ behaves as 
\begin{equation}
\label{GPPZ:bpasym}
  \bp(r) = 2 \sqrt{3}\, \e{-r/l}+\cdots~,
\end{equation}
so that, according to eqn.\ \eqref{bpasym}, $\Delta=\lambda=3$ and
$\hat{\bp}(x)=2\sqrt{3}$. 

In order to study the fluctuation equations, it is advantageous to
introduce the variable
\begin{equation}
\label{GPPZ:vdef}
  v = 1-\e{-2r/l}~.
\end{equation}

The system of equations \eqref{phieqn2} and \eqref{hphieqn2} is
equivalent to a third order ODE for $\varphi$, which is of the
form [\emph{cf.}\ eqn.\ \eqref{eqnfactor}]
\begin{equation} 
\label{GPPZ:3eqn}
  \left[ \frac{d^2}{dv^2} +a(v)\frac{d}{dv} +b(v) \right]
  \left[ \frac{d}{dv} -\frac1{2(1-v)} +\frac{p^2l^2}8
  \right] \varphi(v) =0~.
\end{equation}
The first order factor is obtained from eqn.\ \eqref{chieqn}, and the
coefficients $a(v)$ and $b(v)$ have been found using symbolic
algebra software (Maple),
\begin{align}
\label{GPPZ:a}
 a(v)&= \frac{p^2l^2v^2+8v-2p^2l^2v+36}{v(p^2l^2v^2+4v-p^2l^2v+12)}~, \\
\label{GPPZ:b}
 b(v)&= \frac{p^4l^4v(1-v)^2+ p^2l^2(1-v)(v^2 +4v -16) +4(v^2
 +5v-2)}{4v(p^2l^2v^2+4v-p^2l^2v+12)(1-v)^2}~. 
\end{align}

The solution of the second order ODE,
\begin{equation}
\label{GPPZ:2eqn}
  \left[ \frac{d^2}{dv^2} +a(v)\frac{d}{dv} +b(v) \right] \psi(v) =0~,
\end{equation}
which is regular at $v=0$, is 
\begin{equation}
\label{GPPZ:2sol}
  \psi(v) = C_2 \sqrt{1-v}\left\{ 4(1+v) \,
  \F\left(\frac{3+\alpha}2, \frac{3-\alpha}2;3;v\right) 
  + 8 \,\F\left(\frac{1+\alpha}2,
  \frac{1-\alpha}2;2;v\right) \right\}~,
\end{equation}
where $\alpha=\sqrt{1-p^2l^2}$, and $C_2$ is a constant.

Now, one has to solve the inhomogeneous first order equation
\begin{equation}
\label{GPPZ:order1eqn}
  \left[ \frac{d}{dv} +\frac{5-\alpha^2-v(1-\alpha^2)}{8(1-v)}
  \right] \varphi(v) = \psi(v)~.
\end{equation}
Let us write $\varphi(v)$ as
\begin{equation}
\label{GPPZ:lambdadef}
  \varphi(v) = \sqrt{1-v}\left[ C_1 \e{-(1-\alpha^2)v/8} + 
  \frac{64C_2}{1-\alpha^2}\,
  \F\left(\frac{1+\alpha}2, \frac{1-\alpha}2;2;v\right) + 
  C_2 \lambda(v)
  \right]~, 
\end{equation}
where the term with the constant $C_1$ is the universal solution of
eqn.\ \eqref{GPPZ:order1eqn}. Now, we are left with the simpler
equation for $\lambda(v)$, 
\begin{equation}
\label{GPPZ:lambdaeq}
  \left[ \frac{d}{dv} -\frac{1-\alpha^2}8 \right] \lambda(v)
  = -4(1-v)\,\F\left(\frac{3+\alpha}3,\frac{3-\alpha}2;3;v\right)~,
\end{equation}
but we will not have to solve it explicitly. 

Instead, we shall look at the asymptotic behaviour of the fields at
the horizon. First, let us introduce 
\begin{equation}
\label{GPPZ:udef}
  u = \sqrt{1-v}~,
\end{equation}
so that the horizon is at $u=0$. Then, we can describe the asymptotic
behaviour of the field $\varphi$ by the expansion
\begin{equation}
\label{GPPZ:phiexpand}
  \varphi(u) = u\left[A_1 + \ln u (A_2 u^2 +A_3 u^4 +A_4 u^6 +\cdots)
  + B_1 u^2 +B_2 u^4 + B_3 u^6 +\cdots \right]~.
\end{equation}
This expansion uniquely determines $\varphi$, when the three coefficients
$A_1$, $A_2$ and $B_1$ are specified. The higher order
coefficients can be obtained recursively from the third order
ODE \eqref{GPPZ:3eqn}. We have used Maple to perform
this task. 

The solution \eqref{GPPZ:lambdadef} has only two coefficients, $C_1$
and $C_2$, since we have used only the solution of the second order
ODE \eqref{GPPZ:2eqn} that is regular in the bulk. After
changing variables to $u$, one can see from eqn.\ \eqref{GPPZ:lambdaeq}
that $\lambda(u)$ is of order $u^4 \ln u$. Thus, the coefficients
$A_1$, $A_2$ and $B_1$ are related to $C_1$ and $C_2$ by (see
\cite{Abramowitz} for the asymptotic expansion of the
hypergeometric function)
\begin{align}
\label{GPPZ:A1}
  A_1 &= C_1 + \frac{64 \tilde{C}_2}{1-\alpha^2}~,\\ 
\label{GPPZ:A2}
  A_2 &= 32 \tilde{C}_2~,\\ 
\label{GPPZ:B1}
  B_1 &= \frac{1-\alpha^2}8 C_1 + 16 \tilde{C}_2
  \left[\psi\left(\frac{3+\alpha}2\right) +
  \psi\left(\frac{3-\alpha}2\right) -\psi(1)-\psi(2) \right]~.
\end{align}
Here, we have introduced for brevity of notation
\[ \tilde{C}_2 = C_2 \left[ \Gamma\left(\frac{3+\alpha}2\right)\Gamma\left(\frac{3-\alpha}2\right) \right]^{-1}~.\]
Obviously, $A_1$ is identified with the source, $\hp$, whereas $B_1$ is
the response, $\rp$. Specifying the source $P^i_j \hh^j_i$ determines
also $A_2$ uniquely. In fact, $P^i_j h^j_i$ is obtained from eqn.\
\eqref{normaleqn2}, where $h_{rr}$ is to be expressed in terms of
$\varphi$ by solving algebraically eqns.\ \eqref{phieqn2} and
\eqref{hphieqn2}. After the necessary algebra we find
\begin{equation}
\label{GPPZ:Ph0}
  P^i_j h^j_i(u,p) = \frac{\sqrt{3}}{l^2}[A_1(\alpha^2-1)+2A_2]
  +\mathcal{O}(u^2)~.
\end{equation}
Hence, after determining $A_1$ and $A_2$ in terms of the sources $\hp$
and $P^i_j \hh^j_i$, solving eqns.\ \eqref{GPPZ:A1} and
\eqref{GPPZ:A2} for $C_1$ and $\tilde{C}_2$, and substituting the solution
into eqn.\ \eqref{GPPZ:B1}, we obtain
\begin{equation}
\label{GPPZ:phiresponse}
  \rp(p) = \frac{p^2l^2}4 f_1(p^2) \hp(p) +\frac{\sqrt{3}l^2}{12}
  \left[f_1(p^2)-\frac12\right] P^i_j \hh^j_i(p)~,
\end{equation}
where the function $f_1(p^2)$ is 
\begin{equation}
\label{GPPZ:f1}
  f_1(p^2) = \psi\left(\frac{3+\sqrt{1-p^2l^2}}2\right) +
  \psi\left(\frac{3-\sqrt{1-p^2l^2}}2\right) -\psi(2) -\psi(1)~.
\end{equation}

It remains to determine the response $P^i_j \rh^j_i$. We proceed as
for the derivation of eqn.\ \eqref{GPPZ:Ph0}, but do not truncate at
the lowest order. It is necessary to substitute the recursively
determined coefficients $A_3$, $A_4$, $B_2$ and $B_3$, after which one
finds 
\begin{equation}
\label{GPPZ:Phasym}
\begin{split}
  P^i_j h^j_i(u,p) &=  P^i_j \hh^j_i(p) + u^2  \frac{p^2l^2}8 P^i_j
  \hh^j_i(p) \\
  &\quad + u^4 \left\{\frac{\sqrt{3}p^4l^2}{16} \left[f_1(p^2)
  +\frac12\right] \hp(p) +\frac{p^2l^2}{16} \left[ f_1(p^2) +
  \frac{p^2l^2+4}8 \right] P^i_j \hh^j_i\right\} +\cdots~.
\end{split}
\end{equation}
Here, we have not explicitly written the term of order $u^4 \ln u$, whose
coefficient is a local function of the sources. One can easily check
that eqn.\ \eqref{GPPZ:Phasym} agrees with the general form
\eqref{Phasym} up to terms of order $u^4$ that depend locally on the
sources. The term proportional to $u^4$ is the response $P^i_j \rh^j_i$.

For completeness, we shall repeat here the solution for the traceless
transversal part of $h^i_j$ \cite{Anselmi00a,DeWolfe00a,Bianchi00}. In
terms of the variable $v$, eqn.\ \eqref{tangeqn} reads 
\begin{equation}
\label{GPPZ:htteqn}
  \left[v(1-v)\frac{d^2}{dv^2} + (2-v)\frac{d}{dv} -\frac{p^2l^2}4
  \right] \hTT^i_j(v) = 0~,
\end{equation}
and its solution, which is regular at $v=0$, is 
\begin{equation}
\label{GPPZ:htt}
  \hTT^i_j(v) = C^i_j \,\F\left(\frac{\sqrt{-p^2l^2}}2,
  -\frac{\sqrt{-p^2l^2}}2; 2;v\right)~.
\end{equation}
After asymptotically expanding the hypergeometric function for small
$u=\sqrt{1-v}$ \cite{Abramowitz}, eqn.\ \eqref{GPPZ:htt} can be
expressed as  
\begin{equation}
\label{GPPZ:httasym}
  \hTT^i_j(u) = \hat{\hTT}^i_j(p) \left\{ 1 -\frac{p^2l^2}4 u^2 -
  \frac{p^2l^2(4+p^2l^2)}{32} \left[ 2 \ln u
  +f_2(p^2) +\frac12 \right] u^4
  +\cdots \right\}~,
\end{equation}
where we have introduced the source $\hat{\hTT}^i_j$, and $f_2(p^2)$ is
\begin{equation}
\label{GPPZ:f2}
  f_2(p^2) = \psi\left(2-\frac{\sqrt{-p^2l^2}}2 \right) 
  + \psi\left(2+\frac{\sqrt{-p^2l^2}}2 \right) -2\psi(2)~.
\end{equation}
The term in eqn.\ \eqref{GPPZ:httasym} 
proportional to $u^4$ is the response $\check{\hTT}^i_j$.

We are now in a position to substitute the responses, $\rp$,
$P^i_j\rh^j_i$ and $\check{\hTT}^i_j$, into the expression for the
finite part of the on-shell action, eqn.\ \eqref{Sfin}. One obtains 
\begin{equation}
\label{GPPZ:Sfin}
\begin{split}
  S_{\mathrm{fin}} &= \int\frac{d^4p}{(2\pi)^4} \left\{
  - \hp(-p) \frac{p^2l}4 f_1(p^2) \hp(p) 
  - \hp(-p) \frac{\sqrt{3}l}6 f_1(p^2) P^i_j \hh^j_i(p)\right. \\
  &\quad \left. - \hh^i_j(-p) \frac{l}{4p^2} \left[ \frac13 
  \left(f_1(p^2) +\frac{4+p^2l^2}8 \right) P^j_i P^l_k
  - \frac{4+p^2l^2}8 \left( f_2(p^2) +\frac12 \right) 
  \Pi^{jl}_{ik} \right] \hh^k_l(p) \right\}~.
\end{split}
\end{equation}

Finally, from eqn.\ \eqref{GPPZ:Sfin} one can read off the two-point
functions 
\begin{align}
\label{GPPZ:OO}
  \left\langle \mathcal{O}(p)\mathcal{O}(-p)\right\rangle &= 
  \frac{\kappa l}2 p^2 f_1(p^2)~,\\
\label{GPPZ:TO}
  \left\langle T^i_j(p)\mathcal{O}(-p)\right\rangle &=
  \frac{\sqrt{3}\kappa l}3 f_1(p^2) P^i_j~,\\
\label{GPPZ:TT}
  \left\langle T^i_j(p) T^k_l(-p)\right\rangle &= \frac{2\kappa l}{p^2}
  \left[ -\frac{(4+p^2l^2)}8 f_2(p^2) \Pi^{ik}_{jl} + \frac13 f_1(p^2) P^i_j
  P^k_l \right]~.
\end{align}
In passing from eqn.\ \eqref{GPPZ:Sfin} to eqn.\ \eqref{GPPZ:TT} we
have omitted a local term proportional to $(4+p^2l^2)$.
The correlator \eqref{GPPZ:OO} and the traceless transversal part of
the correlator \eqref{GPPZ:TT} agree with the expressions known from
the literature up to normalization \cite{Anselmi00a,Arutyunov00a,Bianchi00}. 
Taking the trace of the energy momentum tensors, we find the following
relations,
\begin{align}
\label{GPPZ:TO2}
  \left\langle T^i_i(p)\mathcal{O}(-p)\right\rangle &=
  \sqrt{3}\kappa l p^2 f_1(p^2)= 2\sqrt{3}
  \left\langle \mathcal{O}(p)\mathcal{O}(-p)\right\rangle~,\\
\label{GPPZ:TT2}
  \left\langle T^i_i(p) T^j_j(-p)\right\rangle &= 6\kappa l p^2 f_1(p^2)
  =12 \left\langle \mathcal{O}(p)\mathcal{O}(-p)\right\rangle~,
\end{align}
leading to the conclusion that 
\begin{equation}
\label{GPPZ:TOrel}
  T^i_i =2\sqrt{3}\,\mathcal{O}~.
\end{equation}
This relation is direct evidence of the expected operator relation 
\begin{equation}
\label{GPPZ:oprel}
  T^i_i(x) = \beta^n \mathcal{O}_n(x)~,
\end{equation}
where $\beta^n$ are the beta functions corresponding to the couplings
of the primary operators $\mathcal{O}_n$ \cite{Osborn91}. 
For the active scalar operator in the GPPZ flow, we find 
\begin{equation}
\label{GPPZ:beta}
  \beta = 2 \sqrt{3} = -(\Delta -d) \hat{\bp}(x)~,
\end{equation}
which agrees with the renormalized holographic beta function
\cite{Anselmi00a,Kalkkinen01b}.

\subsection{Coulomb Branch Flow}
\label{coulomb}
In this section, we shall consider as an example one of the five
renormalization group flows involving a single active
scalar from the \textbf{20'} of $SO(6)$ in $d=4$, $\mathcal{N}=4$ SYM
theory \cite{Freedman00b,Brandhuber99}. It was called $n=2$ flow in
\cite{Freedman00b}. These flows are vev flows moving the field theory
onto the Coulomb branch.

In our notation, the potential $U(\phi)$ responsible for the Coulomb branch
flow is given by
\begin{equation}
\label{Coulomb:U}
  U(\phi) = -\frac4l \left( \e{-\phi/\sqrt{6}} +\frac12
  \e{2\phi/\sqrt{6}} \right)~.
\end{equation}
Introducing the new variable
\begin{equation}
\label{Coulomb:vdef}
  v = \e{\sqrt{6}\bp(r)/2}~,
\end{equation}
the various quantities can be re-expressed as 
\begin{equation}
\label{Coulomb:rels}
\begin{aligned}
  U(\bp) &= -\frac{2(v+2)}{l v^{1/3}}~, & 
  \frac{\partial U}{\partial \bp} &= \frac{4(1-v)}{\sqrt{6}l
  v^{1/3}}~,\\
  V(\bp) &= \frac{4(1+2v)}{l^2 v^{2/3}}~, & 
  \frac{\partial V}{\partial \bp} &= -\frac{8(1-v)}{\sqrt{6}l^2
  v^{2/3}}~, &
  \frac{\partial^2 V}{\partial \bp^2} &= \frac{4(2+v)}{3l^2 v^{2/3}}~,\\
  \e{2A(r)} &= \frac{l^2 v^{2/3}}{L^2(1-v)}~, &
  \frac{dv}{dr} &= \frac2l v^{2/3} (1-v)~,
\end{aligned}
\end{equation}
where the length $L$ is related to the radius of $D3$--branes, $l_{D3}$,
by $l_{D3} = l^2/L$ \cite{Freedman00b}. 

The horizon is at $v=1$. The asymptotic behaviour of the background
field $\bp$ towards the horizon can be obtained
from eqns.\ \eqref{Coulomb:vdef} and \eqref{Coulomb:rels} (remember
that $A(r)\to r/l$), 
\begin{equation}
\label{Coulomb:bpasym}
  \bp(r) = -\frac{\sqrt{6}l^2}{3L^2} \e{-2r/l}+\cdots~,
\end{equation}
and we can read off $\Delta=2$ and
$\check{\bp}(x)=-\sqrt{6}l^2/(3L^2)$.   

The system of equations \eqref{phieqn2} and \eqref{hphieqn2} is
equivalent to a third order ODE for $\varphi$, which is of the
form 
\begin{equation}
\label{Coulomb:3eqn}
  \left[ \frac{d^2}{dv^2} +a(v)\frac{d}{dv} +b(v) \right]
  \left[ \frac{d}{dv} +\frac1{2+v} +\frac1{1-v}
  +\frac{3L^2p^2}{8v(2+v)}\right] \varphi(v) =0~.
\end{equation}
Again, we have performed the factorization using Maple and obtained
the coefficients $a(v)$ and $b(v)$ as 
\begin{align}
\label{Coulomb:a}
  a(v) &= \frac{5p^2L^2v^2+48v-8p^2L^2+64}{v(2+v)[p^2L^2(v-1)+8]}~,\\
  b(v) &= \frac{p^4L^4(v-1)(v+2) +4p^2L^2(4v^3 -9v^2+8)
  +64(v-1)(3v+2)}{4(v-1)v^2(2+v)[p^2L^2(v-1)+8]}~.
\end{align}

Let us now solve the second order equation,
\begin{equation}
\label{Coulomb:order2eq}
  \left[ \frac{d^2}{dv^2} +a(v)\frac{d}{dv} +b(v) \right] \psi(v) =0~,
\end{equation}
With the substitution 
\begin{equation}
\label{Coulomb:rhodef}
  \psi(v) = (2+v)^{-1} v^{\alpha/2-2} \rho(v)~,
\end{equation}
where the constant $\alpha$ is 
\begin{equation}
\label{Coulomb:alphadef}
  \alpha = \sqrt{1+p^2L^2} +1~,
\end{equation}
eqn.\ \eqref{Coulomb:order2eq} becomes
\begin{multline}
\label{Coulomb:rhoeq}
  v(1-v)\left(v-\frac{(\alpha+2)(\alpha-4)}{\alpha(\alpha-2)} \right) 
  \frac{d^2\rho}{dv^2} \\+
  \left[ -(\alpha-1) v^2 + \frac{(\alpha+2)(\alpha-4)}{\alpha-2} v +
  (\alpha-1) v - \frac{(\alpha+2)(\alpha-4)}{\alpha-2} \right] 
  \frac{d\rho}{dv} \\- 
  \frac14 \left[ (\alpha-2)^2 v - \alpha(\alpha-4) \right] \rho = 0~.
\end{multline}
The solution of eqn.\ \eqref{Coulomb:rhoeq} that is regular at $v=0$ is
\begin{equation}
\label{Coulomb:rhosol}
  \rho(v)= C\left[ \F\left(\frac{\alpha}2-1,\frac{\alpha}2-1;\alpha;v\right) 
  +\frac2\alpha \F\left(\frac{\alpha}2-1,\frac{\alpha}2;\alpha;
  v\right)\right]~,
\end{equation}
where $C$ is a constant.

It remains to extract the asymptotic behaviour of $\varphi(x)$ near
the horizon from the first order differential equation
\begin{equation}
\label{Coulomb:order1eq}
  \left[ \frac{d}{dv} +\frac1{2+v} +\frac1{1-v}
  +\frac{3L^2p^2}{8v(2+v)}\right] \varphi(v) = \psi(v)~.
\end{equation}
Let us introduce the new variable $u$ by $v=1-(l^2/L^2)u$, so that
$u\approx\e{-2r/l}$ for large $r$. Then, the field $\varphi$ has an
asymptotic expansion
\begin{equation}
\label{Coulomb:phiasy}
  \varphi(u) = \ln u \left( A_1 u + A_2 u^2 +A_3 u^3 +\cdots \right)
  + B_1 u + B_2 u^2 +B_3 u^3 +\cdots~.
\end{equation}
Since $\varphi(u)$ satisfies the third order differential equation
\eqref{Coulomb:3eqn}, specifying the three coefficients $A_1$, $B_1$
and $B_2$ determines the field uniquely. The higher order coefficients
can be determined recursively, which we have done using Maple.  
The coefficients $A_1$ and $B_1$ are related to the source and
response, respectively, by
\begin{equation}
\label{Coulomb:ABphi}
  \hp = -2 A_1, \qquad \rp = B_1~.
\end{equation}
In order to determine the coefficient $B_2$, one has to look at the
leading order of the field $P^i_j h^j_i$. It is obtained from eqn.\
\eqref{normaleqn2}, where $h_{rr}$ is to be expressed in terms of
$\varphi$ by solving algebraically eqns.\ \eqref{phieqn2} and
\eqref{hphieqn2}. After the necessary algebra we find
\begin{equation}
\label{Coulomb:Ph0}
  P^i_j h^j_i(u,p) = \frac{\sqrt{6}}{l^2} \left\{ [16-6\alpha(\alpha-2)] A_1
  +[4+3\alpha(\alpha-2)] B_1 +24\frac{L^2}{l^2} B_2 \right\} +
  \mathcal{O}(u)~.
\end{equation}

The additional relation between $A_1$, $B_1$ and $B_2$ is obtained
from eqn.\ \eqref{Coulomb:order1eq} as follows.
After changing the variable to $u$ and
inserting the asymptotic expansion \eqref{Coulomb:phiasy}, the
left hand side of eqn.\ \eqref{Coulomb:order1eq} reads (the
recursively determined higher order coefficients must also be
substituted) 
\begin{multline}
\label{Coulomb:ord1lhs}
  -\frac{L^2}{l^2}A_1 - \frac18 \alpha(\alpha-2) A_1 u \ln u \\
  +\left\{\left[-\frac13-\frac14\alpha(\alpha-2)\right] A_1
  +\left[\frac13+\frac18\alpha(\alpha-2) \right] B_1-
  \frac{L^2}{l^2} B_2\right\} u +\mathcal{O}(u^2\ln u)~,
\end{multline}
whereas the asymptotic expansion of the right hand side is
\cite{Abramowitz} 
\begin{multline}
\label{Coulomb:ord1rhs}
  \frac{2\Gamma(\alpha)C}{3\Gamma(\alpha/2+1)^2} \left\{
  1 + \frac{l^2}{8L^2} \alpha(\alpha-2) u \ln u \right. \\
  \left. +\frac{l^2 u}{8L^2} \left[ \frac{32}8 + \alpha(\alpha-2)
  \left( \ln \frac{l^2}{L^2} +2\psi\left(\frac{\alpha}2\right)
  -2\psi(1) -2 \right) \right] \right\} +\mathcal{O}(u^2\ln u)~.
\end{multline}
Comparing the constant terms of eqns.\ \eqref{Coulomb:ord1lhs} and
\eqref{Coulomb:ord1rhs} yields $C$ in terms of $A_1$, and the $u$ term
gives the desired relation between the three coefficients 
$A_1$, $B_1$ and $B_2$. Expressing $A_1$, $B_1$ and $B_2$ in terms of
the sources $\hp$ and $P^i_j \hh^j_i$ and the response $\rp$, one
finds the following relation,
\begin{equation}
\label{Coulomb:phiresponse}
  \rp(p) = \left[ -\psi \left(\frac{\alpha}2\right) +\psi(1) -\frac12
  \ln \frac{l^2}{L^2} +\frac{4}{3\alpha(\alpha-2)} \right] \hp(p) +
  \frac{\sqrt{6} l^2}{9\alpha(\alpha-2)} P^i_j \hh^j_i(p)~.
\end{equation}

It remains to determine the response $P^i_j \rh^j_i$. We proceed as
for the derivation of eqn.\ \eqref{Coulomb:Ph0}, but do not truncate at
the lowest order. It is necessary to substitute the recursively
determined higher order coefficients, after which one finds 
\begin{equation}
\label{Coulomb:Phasym}
\begin{split}
  P^i_j h^j_i(u,p) &=  P^i_j \hh^j_i(p) + u \frac{p^2l^2}8 P^i_j
  \hh^j_i(p) \\
  &\quad + u^2 \left[\frac{\sqrt{6}p^2l^2}{24L^2}\hp(p)
  +\frac{p^2l^4}{384L^2} (3p^2L^2+32) P^i_j \hh^j_i(p)\right] +\cdots~.
\end{split}
\end{equation}
Again, we have not written explicitely the term $u^2 \ln u$, whose
coefficient depends locally on the sources. The $u^2$ term in eqn.\
\eqref{Coulomb:Phasym} is the response $P^i_j \rh^j_i$.

For completeness, we shall give here also the solution for the
traceless transversal part of the graviton. In terms of $v$, eqn.\
\eqref{tangeqn} reads
\begin{equation}
\label{Coulomb:htteqn}
  \left[ v(1-v) \frac{\partial^2}{\partial v^2} 
  + (2-v) \frac{\partial}{\partial v} -\frac1{4v}\alpha(\alpha-2)
  \right] \hTT^i_j =0~,
\end{equation}
whose solution, which is regular for $v=0$, is 
\begin{equation}
\label{Coulomb:htt}
 \hTT^i_j = C^i_j v^{\alpha/2-1}\, \F\left( \frac{\alpha}2-1,
 \frac{\alpha}2-1;\alpha;v \right)~.
\end{equation}
Asymptotically expanding eqn.\ \eqref{Coulomb:htt} for small $u$
yields \cite{Abramowitz}
\begin{multline}
\label{Coulomb:httasy}
 \hTT^i_j(u,p) = \hat{\hTT}^i_j(p) \left\{ 1 -\frac{p^2l^2}4 u
 \right.\\
 -\left. \frac{p^4l^4}{32} u^2 \left[ \ln u +\ln \frac{l^2}{L^2} +
 2\psi\left(\frac{\alpha}2\right) - 2\psi(1) +\frac{4}{p^2L^2}
 -\frac32 \right] +\cdots\right\}~.   
\end{multline}
The term proportional to $u^2$ is the response $\check{\hTT}^i_j$. 
Notice that, in contrast to the massless scalar field, we cannot
discard the local terms in eqn.\ \eqref{Coulomb:httasy}, because there
is a factor $1/p^4$ in the on-shell action. 

Finally, we substitute the responses $\rp$, $P^i_j\rh^j_i$ and
$\check{\hTT}^i_j$ into the expression for the finite on-shell action,
eqns.\ \eqref{Sfin} and \eqref{d2Sfin}. We obtain
\begin{equation}
\label{Coulomb:Sfin}
\begin{split}
  S_{\mathrm{fin}} &= \int \frac{d^4p}{(2\pi)^4} \left\{ \hp(-p)
  \frac{\sqrt{6}l}{3L^2}\delta(p) + \frac1{2l} \hp(-p) \left[
  \psi\left(\frac{\alpha}2 \right) -\psi(1) +\frac12 \ln
  \frac{l^2}{L^2} - \frac{4}{3p^2L^2} \right] \hp(p) \right. \\
  &\quad -\frac{\sqrt{6}l}{3L^2} \hp(-p) \left(
  \frac{P^i_j}{3p^2}-\frac12 \delta^i_j \right) \hh^j_i(p)
  - \hh^i_j(-p) \frac{l^3}3  P^j_i P^k_l \left( \frac1{32}
  + \frac1{3p^2L^2} \right) \hh^l_k(p) \\
  &\quad \left. + \hh^i_j(-p) \frac{l^3}{16}\Pi^{jk}_{il}
  \left[ \psi\left(\frac{\alpha}2 \right) -\psi(1) +\frac12 \ln
  \frac{l^2}{L^2} +\frac{2}{p^2L^2} -\frac34 \right]
  \hh^l_k(p) \right\}~.
\end{split}
\end{equation}
From eqn.\ \eqref{Coulomb:Sfin} we can read off the
correlators  
\begin{align}
\label{Coulomb:O}
  \langle\mathcal{O}(p) \rangle &= \frac{\sqrt{6}\kappa l}{3L^2}
  \delta(p)~,\\ 
\label{Coulomb:OO}
  \langle\mathcal{O}(p)\mathcal{O}(q) \rangle -
  \langle\mathcal{O}(p)\rangle\langle \mathcal{O}(q) \rangle &= 
  -\frac{\kappa}l \left[
  \psi\left(\frac{\alpha}2 \right) -\psi(1) - \frac{4}{3p^2L^2}
  \right] \delta(p+q)~,\\
\label{Coulomb:TO}
  \left\langle T^i_j(-p)\mathcal{O}(p) \right\rangle &=
  \frac{\sqrt{6}\kappa l}{3L^2} \left( \frac2{3p^2} P^i_j-\delta^i_j
  \right)~,\\ 
\label{Coulomb:TT}
  \left\langle T^i_j(-p) T^k_l(p) \right\rangle &=
  - \frac{\kappa l^3}{2}\Pi^{ik}_{jl}
  \left[ \psi\left(\frac{\alpha}2 \right) -\psi(1) -\frac{2}{3p^2L^2}
  \right]~,  
\end{align}
where we have discarded some local terms. In particular, we have
shifted the $p^2=0$ pole term proportional to $P^i_jP^k_l$ in the
energy momentum tensor two-point function to the traceless transversal
part. Notice that the term propotional to $\delta^i_j$ in
$\left\langle T^i_j(-p)\mathcal{O}(p) \right\rangle$ has its origin in
the source term inserted into the boundary field theory action
functional. Thus, $T^i_j$ is not exactly the energy momentum operator
of the field theory, and, when one considers the true operator, this
unusal term does not appear \cite{Bianchi01}. 
 Eqn.\ \eqref{Coulomb:OO} has the
opposite sign with respect to the result of DeWolfe and Freedman
\cite{DeWolfe00a}. This is a result of adding the counter term
\eqref{Sct}, which contributes to the finite on-shell action. For
large $|p|$, eqn.\ \eqref{Coulomb:OO} approaches the conformal
two-point function for operators of dimension $\Delta=2$ (with the
correct sign), so that we conclude that our sign is the correct one.
The pole at $p^2=0$ can be interpreted by the existence of a massless 
dilaton from the broken conformal symmetry. The traceless transversal part
of eqn.\ \eqref{Coulomb:TT} has already been found in
\cite{Freedman00b}.

\section{Conclusions}
\label{conc}
In this paper, we have presented a detailed calculation of one- and
two-point correlation functions in holographic renormalization group
flows dual to the 
Poincar\'e invariant background solutions of generic bulk theories
containing scalars coupled to gravity. 
We were able to explicitly distinguish between operator and vev flows
by means of the condensate in the boundary field theory. This result
was achieved by carefully considering the counter terms, which must be
added to the bulk field theory action in order to cancel divergences. 
For operators of dimension $\Delta>d/2$, we can describe both operator
and vev flows, whereas for operators of dimension $\Delta=d/2$ we only
have vev flows. 

Our main improvement for two-point functions in holographic
renormalization group flows  is that we considered the
sources of the metric and scalar fluctuations as independent boundary
data, which couple as sources to the energy momentum tensor and scalar
operators of the boundary field theory, respectively. Thus, our method
results in the complete non-local correlation functions $\langle
\mathcal{O}\mathcal{O}\rangle$, $\langle T^i_j\mathcal{O}\rangle$ and $\langle
T^i_j T^k_l\rangle$. An analysis of these correlation functions yields
important facts about the boundary field theory. For example, our
analysis of the GPPZ flow in Sec.\ \ref{GPPZ} confirmed the operator
relation $T^i_i = \beta\mathcal{O}$, where $\beta$ is the renormalized
holographic beta function. Moreover, there should be a relation
between the $P^i_j P^k_l$ part of $\langle T^i_j T^k_l\rangle$ and the
$c$-function of the RG flow \cite{Anselmi99}. In order to obtain the
correct contact terms in the correlators, one would have to carefully
analyze \emph{all} divergent terms in the on-shell action and to add
necessary counter terms along the lines of \cite{deHaro00a}. This has
been achieved very recently in \cite{Bianchi01}.

We have chosen to apply our improved method to the GPPZ and Coulomb
branch flows, which have already been studied in the literature, 
in order to be able to directly verify the known results and to
improve them by finding the complete set of two-point functions.  
We leave the application of our method to other holographic
renormalization group flows for the future. In particular, a flow
that interpolates between UV and IR conformal fixed points would be
specially interesting, because the correlation functions should
exhibit this flow in their dependence on the momentum $p$. In that
respect, toy models, which do not necessarily contain the super
potentials of true super gravity theories, might be very
valuable. Furthermore, our method remains to be tested for a vev flow
with $\Delta>d/2$.

Finally, we would like to mention several interesing questions. 
It remains an open problem how to describe
holographically RG flows generated by operators of dimension $\Delta$ with
$d/2-1\le\Delta<d/2$ ($d/2-1$ is the unitarity bound). For the AdS
background, the method of using the Legendre transformed action as the
generating functional \cite{Klebanov99-1,Mueck99-4} yields the
correct conformal correlation functions, but it is not yet known
whether it can be generalized to RG flows. One might expect that a
background solution describing an operator flow with $d/2<\Delta<d/2+1$ also
describes a vev flow with a condensate of an operator of dimension
$\Delta'=d-\Delta$. A technical problem is that the known
examples of RG flows do not involve scalar operators with dimensions 
within this interval. 

Another interesing question is to find ``conjugate'' flows and the
relations of their field theoretical duals. By ``conjugate'' we mean
the following. Whether a background solution describes an operator or a
vev flow is determined by the super potential $U$, because of eqn.\
\eqref{bg}. Given a bulk potential $V$, the super potential is not
unique, but there are several solutions. According to eqn.\ \eqref{V},
one could change the overall sign of $U$, but this corresponds to
simply changing the sign of $r$, so that $r\to -\infty$ becomes the
asymptotically AdS region, and no new physics appears. It is more
interesting to consider the expansion around the fixed point, eqn.\
\eqref{Uasym}. In general, there are two solutions for $\lambda$,
$\lambda>d/2$ describing an operator flow, and $\lambda<d/2$
describing a vev flow. We call these two flows ``conjugate,'' and we 
conjecture that the corresponding boundary field theories are
related to each other.

\begin{ack}
I am thankful to M.~Bianchi, D.Z.~Freedman and K.~Skenderis for e-mail
correspondence and their critical comments on the first version of
this paper. In a very recent paper, they have considered the same
problems and obtained results essentially identical with mine
\cite{Bianchi01}. Their approach includes all counter terms, and thus 
yields the correct contact terms in the correlators, with which 
I was not very careful.
Moreover, I would like to thank J.~Erdmenger, A.~Petkou,
R.~Pettorino and M.~Porrati for useful discussions. 
Financial support by the European Commission RTN programme 
HPRN-CT-2000-00131, in which I am associated with INFN, Sezione di
Frascati, is gratefully acknowledged.  
\end{ack}

\begin{appendix}
\section{Geometric Relations for Hyper Surfaces}
\label{geom}
This appendix shall begin with a review of basic geometric relations
for hyper surfaces \cite{Eisenhart}. A hyper surface in a
space--time with coordinates $X^\mu$ ($\mu=0,\ldots, d$) is defined as
a set of $d+1$ functions, $X^\mu(x^i)$ ($i=1,\ldots, d$), where $x^i$
are the coordinates of the hyper surface. Moreover, let $\tg_{\mu\nu}$
and $g_{ij}$ be the metric tensors of 
space--time and the hyper surface, respectively. The
tangents $X^\mu_i \equiv \partial_i X^\mu$ and the normal $N^\mu$ of
the hyper surface satisfy the following orthogonality relations,
\begin{align}
\label{geom:induced}
  \tilde g_{\mu\nu}\, X^\mu_i X^\nu_j &= g_{ij}~,\\
\label{geom:orth}
  X^\mu_i N_\mu &= 0~,\\
\label{geom:norm}
  N^\mu N_\mu &=1~.
\end{align}
As in the main text, a tilde will be used to label quantities
characterizing the ($d+1$)--dimensional space--time manifold, whereas
those of the hyper surface remain unadorned. 

The equations of Gauss and Weingarten define the second fundamental
form, $H_{ij}$, of the hyper surface,
\begin{align}
\label{geom:gauss1}
  \partial_i X^\mu_j + \tG^\mu{}\!_{\lambda\nu} X^\lambda_i X^\nu_j
  - \Gamma^k{}\!_{ij} X^\mu_k = H_{ij} N^\mu~,\\
\label{geom:weingarten}
  \partial_i N^\mu + \tG^\mu{}\!_{\lambda\nu} X^\lambda_i N^\nu 
  = -H^j_i X^\mu_j~.
\end{align}
The second fundamental form describes the extrinsic curvature of the
hyper surface and is related to the intrinsic curvature by another
equation of Gauss,
\begin{align}
\label{geom:gauss2}
  \tilde R_{\mu\nu\lambda\rho} X^\mu_i X^\nu_j X^\lambda_k X^\rho_l 
  &= R_{ijkl} + H_{il} H_{jk} - H_{ik} H_{jl}~.\\
\intertext{Furthermore, it satisfies the equation of Codazzi,}
\label{geom:codazzi}
  \tilde R_{\mu\nu\lambda\rho} X^\mu_i X^\nu_j N^\lambda X^\rho_k 
  &= \nabla_i H_{jk} - \nabla_j H_{ik}~. 
\end{align}

The above formulae simplify, if (as in the familiar time slicing
formalism \cite{Wald,MTW}), we choose space--time coordinates such
that 
\begin{equation}
\label{geom:Xdef}
  X^0 = r~, \quad X^i=x^i~.
\end{equation}
Then, the tangent vectors are given by $X^0_i=0$ and $X^j_i
=\delta_i^j$. One conveniently splits up the space--time
metric as (shown here for Euclidean signature)
\begin{align}
\label{geom:split}
  \tg_{\mu\nu} &= \begin{pmatrix} n_i n^i +n^2 & n_j \\
				n_i & g_{ij} \end{pmatrix}~,\\
\intertext{whose inverse is given by}
\label{geom:splitinv}
  \tg^{\mu\nu} &= \frac1{n^2} \begin{pmatrix} 1& -n^j \\
			-n^i & n^2 g^{ij} +n^i n^j \end{pmatrix}~.
\end{align}
The matrix $g^{ij}$ is the inverse of $g_{ij}$ and is used to raise
hyper surface indices. The quantities $n$ and
$n^i$ are the lapse function and shift vector, respectively. 

The normal vector $N^\mu$ satisfying the eqns.\ 
\eqref{geom:orth} and \eqref{geom:norm} is given by 
\begin{equation}
\label{geom:normal}
  N_\mu = (n,0)~, \qquad N^\mu= \frac1{n}(1,-n^i)~.
\end{equation}
Then, one obtains the second fundamental form from the equation of
Gauss, eqn.\ \eqref{geom:gauss1}, as 
\begin{equation}
\label{sli:Hij}
  H_{ij} = - \frac1{2n} \left(\partial_r g_{ij} -
  \nabla_i n_j - \nabla_j n_i \right)~.
\end{equation}

Finally, using the equations of Gauss and Weingarten, certain
components of the space--time connections can be expressed as follows,
\begin{align}
\label{geom:conn1}
  \tG^r{}\!_{ij} &= \frac1n H_{ij}~,\\
\label{geom:conn2}
  \tG^k{}\!_{ij} &= \Gamma^k{}\!_{ij} - \frac{n^k}n H_{ij}~,\\
\label{geom:conn3}
  \tG^r{}\!_{ir} &= \frac1n \partial_i n + \frac{n^j}n H_{ij}~,\\
\label{geom:conn4}
  \tG^k{}\!_{ir} &= \nabla_i n^k - \frac{n^k}n \partial_i n 
  - nH_{ij} \left( g^{jk}+ \frac{n^jn^k}{n^2} \right)~.
\end{align}
\end{appendix}

\end{document}